\documentclass{Interspeech}

\interspeechcameraready

\title{Attention Is Not Always the Answer: Optimizing Voice Activity Detection with Simple Feature Fusion}

\author[equalcontribution]{Kumud}{Tripathi}
\author[equalcontribution]{Chowdam Venkata}{Kumar}
\author[]{Pankaj}{Wasnik}

\affiliation[nocounter]{}{Media Analysis Group}{Sony Research India}
\email{\{kumud.tripathi, chowdam.kumar, pankaj.wasnik\}@sony.com}

\usepackage{comment}
\usepackage{multirow}
\usepackage{float}
 
\begin{document}

\maketitle

\begin{abstract}

    Voice Activity Detection (VAD) plays a key role in speech processing, often utilizing hand-crafted or neural features. This study examines the effectiveness of Mel-Frequency Cepstral Coefficients (MFCCs) and pre-trained model (PTM) features, including wav2vec 2.0, HuBERT, WavLM, UniSpeech, MMS, and Whisper. We propose FusionVAD, a unified framework that combines both feature types using three fusion strategies: concatenation, addition, and cross-attention (CA). Experimental results reveal that simple fusion techniques, particularly addition, outperform CA in both accuracy and efficiency. Fusion-based models consistently surpass single-feature models, highlighting the complementary nature of MFCCs and PTM features. Notably, our best-performing fusion model exceeds the state-of-the-art Pyannote across multiple datasets, achieving an absolute average improvement of 2.04\%. These results confirm that simple feature fusion enhances VAD robustness while maintaining computational efficiency.
    
\end{abstract}

\section{Introduction}

    



Voice Activity Detection (VAD) is the task of detecting speech segments within an audio signal \cite{vad}. It serves as a fundamental pre-processing step for various speech-related applications, including Automatic Speech Recognition (ASR), Speaker Recognition, Speaker Verification, and Speaker Diarization \cite{asr, sid, sd}. By accurately identifying speech and non-speech regions, VAD significantly enhances the performance of these systems by filtering out non-speech and noisy segments. As these speech-based technologies become more prevalent in applications such as virtual assistants, hearing aids, and telecommunications, improving VAD accuracy has become crucial for enhancing both user experience and system robustness. The problem of VAD has been an active research topic for several decades, typically approached as a frame-level classification task, distinguishing between speech and non-speech. Traditional approaches to VAD employ threshold-based or statistical machine learning methods using acoustic features such as energy, zero-crossing rate, pitch, and auto-correlation \cite{patil2024unveiling, alimi2022voice, nguyen2021analysis}. While, these methods perform well in clean environments, they often fail in real-world scenarios where background noise and varying acoustic conditions degrade their reliability.


Modern deep learning approaches for VAD, including Convolutional Neural Networks (CNNs) \cite{cnn} and Recurrent Neural Networks (RNNs) \cite{rnn}, have demonstrated superior performance by effectively integrating frequency-domain filtering with temporal sequence modeling. These architectures enhance the robustness of VAD models in real-world noisy environments by jointly learning feature extraction and task modeling \cite{wilkinson2021hybrid, zazo2016feature}. However, their performance heavily relies on the availability of large-scale labeled datasets. In contrast, pre-trained models (PTMs) such as wav2vec 2.0 \cite{baevski2020wav2vec}, HuBERT \cite{hsu2021hubert}, and WavLM \cite{chen2022wavlm} utilize vast amounts of unlabeled speech data to learn generalized representations using CNNs and Transformers. 
 wav2vec 2.0 and HuBERT efficiently capture phonetic structures, while WavLM enhances robustness \cite{shankar2024closer}. UniSpeech and Massively Multilingual Speech (MMS) leverage multilingual learning \cite{chen2022unispeech,mms},  
while Whisper demonstrates strong performance in large-scale ASR tasks \cite{radford2023robust}, including those involving low-resource languages \cite{kumud}. Their diverse learning paradigms provide valuable insights for downstream speech processing tasks \cite{mohamed2022self}.

 These PTM models have demonstrated success in several binary classification tasks, including DeepFake detection \cite{phukan2024investigating} and Violence Detection  \cite{phukan2024multi}. Given that VAD is also a binary classification task (Speech vs. Non-Speech), PTM-based approaches are particularly well-suited for this problem. Previous studies have employed PTM models for VAD by fine-tuning them on task-specific labeled datasets, demonstrating state-of-the-art results \cite{karan2024transformer, kunevsova2023multitask}. However, there has been limited exploration of why PTM features perform well for VAD and how they compare to traditional hand-crafted features like MFCC \cite{mfcc}. Additionally, the potential benefits of combining PTM features with MFCC for VAD remain largely unexplored. 

\begin{figure*}
    \centering
    \includegraphics[width=0.9\linewidth]{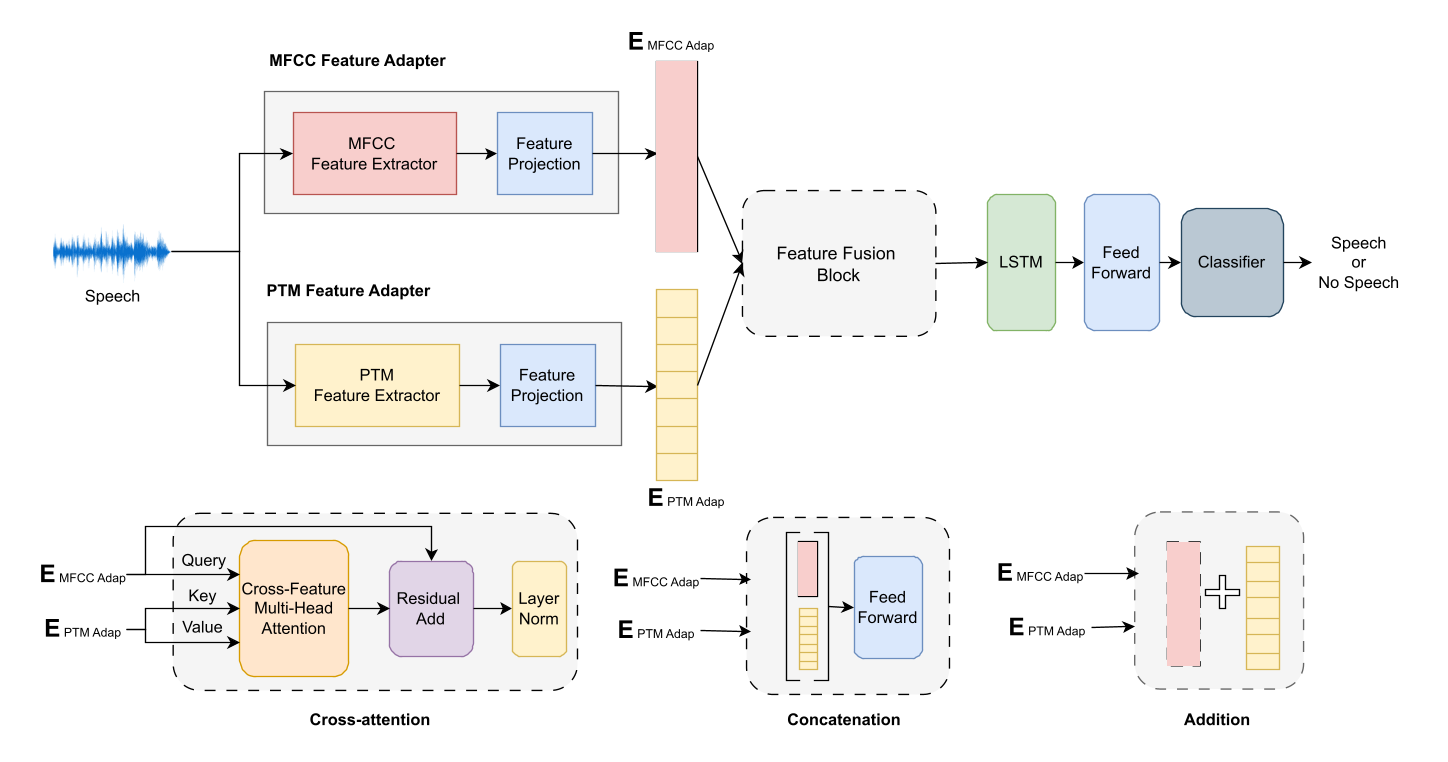}
    \vspace{-4mm}
    \caption{Overview of the FusionVAD Framework with Different Feature Fusion Strategies.}
    \label{fig:block}
    \vspace{-4mm}
\end{figure*}

In this work, we systematically analyze the effectiveness of MFCCs and PTM-based speech representations for VAD. 
We explore wav2vec 2.0, HuBERT, WavLM, UniSpeech, MMS, and Whisper as PTMs due to their proven success in various speech-processing tasks. First, we compare the performance of VAD models trained separately with MFCCs and PTM representations and analyze their respective failure cases. Next, we explore different feature fusion techniques, including concatenation, addition, and cross-attention, to combine MFCC and PTM representations. Our experiments on publicly available datasets such as AMI \cite{ami}, Callhome \cite{Callhome}, and VoxConverse \cite{VoxConverse} reveal that both MFCC and PTM features contain complementary information, which, when effectively fused, enhances VAD performance. Surprisingly, we find that simple fusion techniques like concatenation and addition outperform cross-attention-based fusion, challenging the common assumption that complex attention mechanisms are always necessary for effective speech classification.
The key contributions of this work are as follows:
\begin{enumerate}
    \item Examines the role of attention mechanism in feature fusion for Voice Activity Detection (VAD) and shows that attention-based fusion is not always necessary for effective speech and non-speech classification.
    \item Introduces a simple yet effective feature fusion method that combines MFCC and PTM representations.
    \item Conducts a comprehensive analysis of state-of-the-art PTMs to evaluate their effectiveness for VAD.
    \item Demonstrates that addition-based feature fusion enhances both accuracy and computational efficiency.
\end{enumerate}





\section{Methodology}

\subsection{MFCC vs PTM Features}


Pre-trained model based features have proven effective for various speech tasks, including VAD. These models leverage self-attention to capture long-range dependencies and generate contextual representations, which are particularly beneficial in noisy environments. In contrast, traditional hand-crafted features like spectrograms and MFCCs offer static time-frequency representations. While such information may be sufficient for speech detection in clean conditions, it becomes less effective in noisy settings where overlapping frequency components obscure speech cues. In these cases, PTM features provide more robust representations, as they are trained on diverse acoustic conditions and noise types. This robustness makes them valuable for improving VAD performance under challenging conditions. Understanding the strengths of PTM features relative to traditional features is essential, especially if they are found to encode complementary information. Combining both types of features could potentially enhance overall VAD performance by leveraging the contextual awareness of PTMs and the fine-grained spectral detail of hand-crafted features.

\begin{table*}[]
\centering
\caption{Performance (in \%) of Voice activity detection with and without feature fusion. *Bold represents the best result.}
\vspace{-2mm}
\label{tab:der}
\begin{tabular}{|c|ccc|ccccccccc|}
\hline
\multirow{3}{*}{\textbf{Feature Extractor}} & \multicolumn{3}{c|}{\multirow{2}{*}{\textbf{Base}}}                                & \multicolumn{9}{c|}{\textbf{Feature   Fusion}}                                                                                                                                                                                                                                                            \\ \cline{5-13} 
                                            & \multicolumn{3}{c|}{}                                                              & \multicolumn{3}{c|}{\textbf{Concatenation}}                                                               & \multicolumn{3}{c|}{\textbf{Addition}}                                                                   & \multicolumn{3}{c|}{\textbf{Cross-attention}}                                      \\ \cline{2-13} 
                                            & \multicolumn{1}{c|}{\textbf{DER}} & \multicolumn{1}{c|}{\textbf{FAR}} & \textbf{MR} & \multicolumn{1}{c|}{\textbf{DER}}   & \multicolumn{1}{c|}{\textbf{FAR}} & \multicolumn{1}{c|}{\textbf{MR}} & \multicolumn{1}{c|}{\textbf{DER}}  & \multicolumn{1}{c|}{\textbf{FAR}} & \multicolumn{1}{c|}{\textbf{MR}} & \multicolumn{1}{c|}{\textbf{DER}} & \multicolumn{1}{c|}{\textbf{FAR}} & \textbf{MR} \\ \hline
\textbf{MFCC}                               & \multicolumn{1}{c|}{6.79}         & \multicolumn{1}{c|}{3.23}        & 3.56        & \multicolumn{1}{c|}{-}              & \multicolumn{1}{c|}{-}           & \multicolumn{1}{c|}{-}           & \multicolumn{1}{c|}{-}             & \multicolumn{1}{c|}{-}           & \multicolumn{1}{c|}{-}           & \multicolumn{1}{c|}{-}            & \multicolumn{1}{c|}{-}           & -           \\ \hline
\textbf{wav2vec 2.0}                           & \multicolumn{1}{c|}{6.70}         & \multicolumn{1}{c|}{1.77}        & 4.92        & \multicolumn{1}{c|}{5.95}           & \multicolumn{1}{c|}{1.82}        & \multicolumn{1}{c|}{4.14}        & \multicolumn{1}{c|}{6.74}          & \multicolumn{1}{c|}{1.96}        & \multicolumn{1}{c|}{4.78}        & \multicolumn{1}{c|}{6.12}         & \multicolumn{1}{c|}{2.08}        & 4.04        \\ \hline
\textbf{HuBERT}                             & \multicolumn{1}{c|}{7.51}         & \multicolumn{1}{c|}{1.80}        & 5.71        & \multicolumn{1}{c|}{4.95}           & \multicolumn{1}{c|}{2.29}        & \multicolumn{1}{c|}{2.66}        & \multicolumn{1}{c|}{6.82}          & \multicolumn{1}{c|}{0.92}        & \multicolumn{1}{c|}{5.89}        & \multicolumn{1}{c|}{6.39}         & \multicolumn{1}{c|}{2.01}        & 4.38        \\ \hline
\textbf{WavLM}                              & \multicolumn{1}{c|}{6.05}         & \multicolumn{1}{c|}{2.03}        & 4.01        & \multicolumn{1}{c|}{5.42}           & \multicolumn{1}{c|}{1.94}        & \multicolumn{1}{c|}{3.48}        & \multicolumn{1}{c|}{4.95}          & \multicolumn{1}{c|}{2.66}        & \multicolumn{1}{c|}{2.30}        & \multicolumn{1}{c|}{5.81}         & \multicolumn{1}{c|}{3.67}        & 2.14        \\ \hline
\textbf{UniSpeech}                          & \multicolumn{1}{c|}{6.25}         & \multicolumn{1}{c|}{2.22}        & 4.03        & \multicolumn{1}{c|}{5.58}           & \multicolumn{1}{c|}{2.19}        & \multicolumn{1}{c|}{3.38}        & \multicolumn{1}{c|}{5.55}          & \multicolumn{1}{c|}{3.34}        & \multicolumn{1}{c|}{2.21}        & \multicolumn{1}{c|}{6.44}         & \multicolumn{1}{c|}{2.26}        & 4.18        \\ \hline
\textbf{MMS}                                & \multicolumn{1}{c|}{6.33}         & \multicolumn{1}{c|}{1.45}        & 4.88        & \multicolumn{1}{c|}{5.11}           & \multicolumn{1}{c|}{2.91}        & \multicolumn{1}{c|}{2.19}        & \multicolumn{1}{c|}{4.87}          & \multicolumn{1}{c|}{2.32}        & \multicolumn{1}{c|}{2.55}        & \multicolumn{1}{c|}{5.93}         & \multicolumn{1}{c|}{3.38}        & 2.55        \\ \hline
\textbf{Whisper}                            & \multicolumn{1}{c|}{5.83}         & \multicolumn{1}{c|}{2.55}        & 3.28        & \multicolumn{1}{c|}{4.70}           & \multicolumn{1}{c|}{1.61}        & \multicolumn{1}{c|}{3.09}        & \multicolumn{1}{c|}{\textbf{4.50}} & \multicolumn{1}{c|}{1.74}        & \multicolumn{1}{c|}{2.76}        & \multicolumn{1}{c|}{5.34}         & \multicolumn{1}{c|}{2.88}        & 2.46        \\ \hline
\end{tabular}
\end{table*}

\begin{table*}[]
\centering
\caption{Comparison (in \%) of best performing fusion model with baseline Pyannote. }
\label{tab:pyannet-whisper}
\vspace{-2mm}
\begin{tabular}{|c|ccc|ccc|ccc|}
\hline
\multirow{2}{*}{\textbf{Model}} & \multicolumn{3}{c|}{\textbf{AMI}}                                                       & \multicolumn{3}{c|}{\textbf{Callhome}}                                               & \multicolumn{3}{c|}{\textbf{VoxConverse}}                                                       \\ \cline{2-10} 
                                & \multicolumn{1}{c|}{\textbf{DER}}  & \multicolumn{1}{c|}{\textbf{FAR}}   & \textbf{MR}   & \multicolumn{1}{c|}{\textbf{DER}}  & \multicolumn{1}{c|}{\textbf{FAR}}   & \textbf{MR}   & \multicolumn{1}{c|}{\textbf{DER}}  & \multicolumn{1}{c|}{\textbf{FAR}}   & \textbf{MR}   \\ \hline
\textbf{Pyannote \cite{bredin2020pyannote}}                & \multicolumn{1}{c|}{11.07}         & \multicolumn{1}{c|}{\textbf{1.70}} & 9.37          & \multicolumn{1}{c|}{4.68}          & \multicolumn{1}{c|}{\textbf{0.54}} & 4.14          & \multicolumn{1}{c|}{3.89}          & \multicolumn{1}{c|}{2.40}          & 1.49          \\ \hline
\textbf{Whisper-MFCC-Addition}       & \multicolumn{1}{c|}{\textbf{7.25}} & \multicolumn{1}{c|}{2.73}          & \textbf{4.53} & \multicolumn{1}{c|}{\textbf{3.28}} & \multicolumn{1}{c|}{0.67}          & \textbf{2.61} & \multicolumn{1}{c|}{\textbf{2.97}} & \multicolumn{1}{c|}{\textbf{1.82}} & \textbf{1.15} \\ \hline
\end{tabular}
\vspace{-4mm}
\end{table*}

To analyze different feature types for VAD, we train models using both hand-crafted and PTM features separately. MFCCs represent hand-crafted features, while PTM features include speech encoders such as wav2vec 2.0, HuBERT, WavLM, UniSpeech, MMS, and Whisper. All speech encoders remain frozen during training, focusing on evaluating their effectiveness for VAD rather than fine-tuning them.
For a fair comparison, we use the same architecture across all models. This architecture, referred to as FusionVAD, replaces the feature fusion block with a feedforward layer, allowing the use of one feature type at a time. The features first pass through two fully connected layers with a hidden size of 128 and GELU activation. Then, two bidirectional LSTM layers (hidden size 128) capture sequence information, which enhances VAD robustness against noise. Finally, two linear layers (hidden size 128, GELU activation) and a classification layer with sigmoid activation produce the output.
Additionally, we visualize model predictions to identify failure cases and highlight the complementary nature of hand-crafted and PTM features, further reinforcing the importance of feature selection for improving VAD performance.


\subsection{Feature Fusion Techniques}
 To integrate MFCC features with PTM representations, we explore three feature fusion techniques: concatenation, addition, and cross-attention. The block diagram (Figure \ref{fig:block}) illustrates the FusionVAD pipeline, where both MFCC and PTM features are extracted, projected, and fused before being processed by an LSTM, followed by a feedforward network and classifier to determine speech or non-speech.
The overall architecture remains consistent across all fusion methods, with differences only in the feature fusion block. Initially, MFCC and PTM features are passed through feature projection layers—fully connected layers that map both feature sets to a 128-dimensional space. Each fusion technique is implemented as follows:
\begin{enumerate}
    \item \textbf{Concatenation:} The MFCC and PTM features are concatenated along the feature dimension and then passed through a fully connected layer to project them back to 128 dimensions before being input to the LSTM.
    \item \textbf{Addition:} Element-wise addition is performed between the MFCC and PTM feature vectors, directly combining their information.
    \item \textbf{Cross-Attention:} Projected MFCC features serve as queries, while the {PTM} features act as keys and values in a multi-head attention mechanism with a 128-dimensional hidden space and two attention heads. A residual connection adds the original projected MFCC features to the cross-attended output to retain important spectral information, followed by layer normalization for stable feature representation.
\end{enumerate}
These fusion techniques aim to combine complementary information from spectral and learned representations to improve VAD performance.

\section{Experiments}\label{sec3}
\subsection{Dataset and Evaluation Metrics}

We conducted all our experiments on three publicly available datasets, i.e., AMI, Callhome, and VoxConverse, to ensure domain diversity. 
We followed the dataset split methodology from \cite{bredin2021end}. Since VoxConverse lacks an official training set, we partitioned its development set into 144 training files and 72 development files. For the Callhome dataset, comprising unscripted English telephone conversations, we selected 139 files: 89 for training, and 25 each for development and testing, resulting in 22 hours of training data. The AMI Corpus was used with its official split, but we restricted training to the first 10 minutes of each file to maintain consistency in training durations across datasets, yielding 22 hours of training data. In total, we used approximately 75.4 hours of audio: VoxConverse contributed 19 hours (15 training, 2 development, 2 testing), AMI contributed 26 hours (22 training, 2 development, 2 testing), and Callhome contributed 30.4 hours (22 training, 4.2 development, 4.2 testing). Performance evaluation was conducted using standard VAD metrics: False Alarm Rate (FAR), Missing Rate (MR), and Detection Error Rate (DER), where DER is the sum of FAR and MR. These metrics provide a comprehensive measure of VAD performance across different datasets.

 \subsection{Experimental Setup}

 Training is configured for 50 epochs for all model training. Early stopping criteria with 5 epoch patience is used to avoid over fitting.  Area under ROC on validation dataset is used for early stopping and also to select the best checkpoints. Models are trained on 2 seconds of chunks with a batch size of 32. All features are extracted with a stride of 20 ms. Pyannote toolkit \cite{bredin2020pyannote} is used for training and testing the models and PTM speech encoders model checkpoints are obtained from huggingface. Base version checkpoints are considered for wav2vec 2.0\footnote{\url{https://huggingface.co/facebook/wav2vec2-base}}, HuBERT\footnote{\url{https://huggingface.co/facebook/hubert-base-ls960}}, WavLM\footnote{\url{https://huggingface.co/patrickvonplaten/wavlm-libri-clean-100h-base-plus}}, UniSpeech \footnote{\url{https://huggingface.co/microsoft/unispeech-sat-base-100h-libri-ft}}and Whisper\footnote{\url{https://huggingface.co/openai/whisper-base}}. One billion parameters checkpoints is used for MMS\footnote{\url{https://huggingface.co/facebook/mms-1b-all}}. We consider baseline as Pyannote VAD \cite{bredin2021end} (official implementation in Pyannote toolkit is used for training) to compare with best performing model. Pyannote VAD follows same architecture except the initial feature extractor, which is Sincnet \cite{sincnet}. We train the Pyannote VAD with the same datasets used for other models. 
 
\begin{figure}[ht]
    \centering
    \includegraphics[width=1\linewidth]{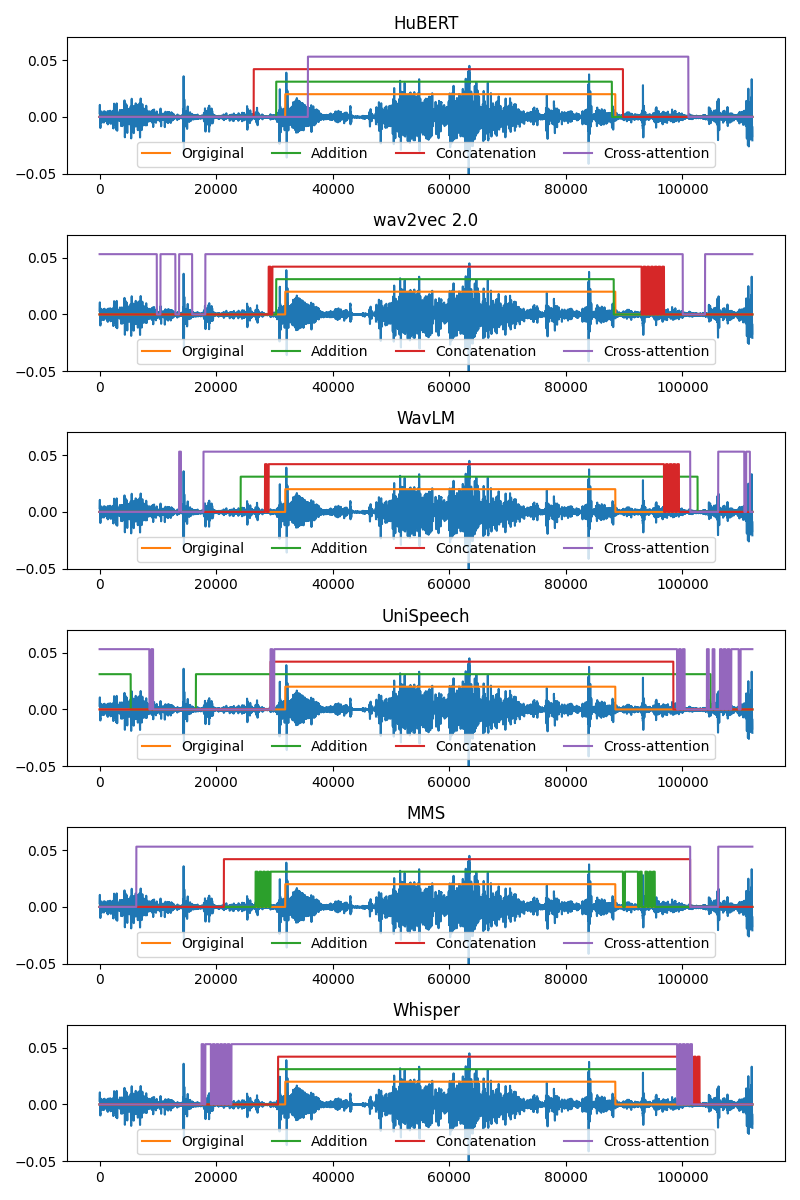}
    \vspace{-6mm}
    \caption{Feature fusion outputs (Green: Addition, Red: Concatenation, and Purple: Cross-Attention) along with the original reference (Yellow) for all FusionVAD models on a single audio segment from the AMI file "EN2004a".}
    \label{fig:fusion-comparison}
    \vspace{-2mm}
\end{figure}

\begin{figure}[ht]
    \centering
    \includegraphics[width=0.9\linewidth]{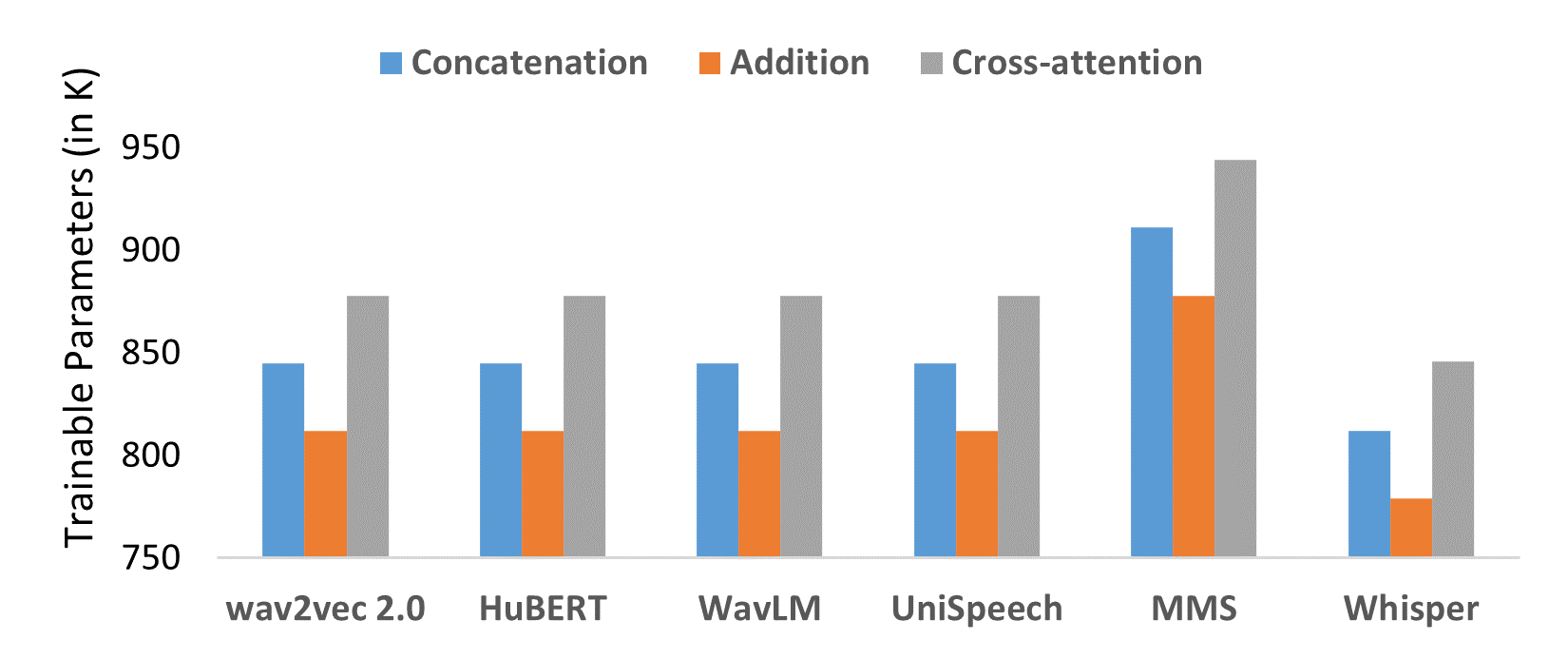}
    \vspace{-2mm}
    \caption{Analysis of trainable parameters for all the FusionVAD models using different fusion techniques.}
    \label{fig:parameters}
    \vspace{-4mm}
\end{figure}

\begin{figure}[ht]
    \centering
    \includegraphics[width=1\linewidth]{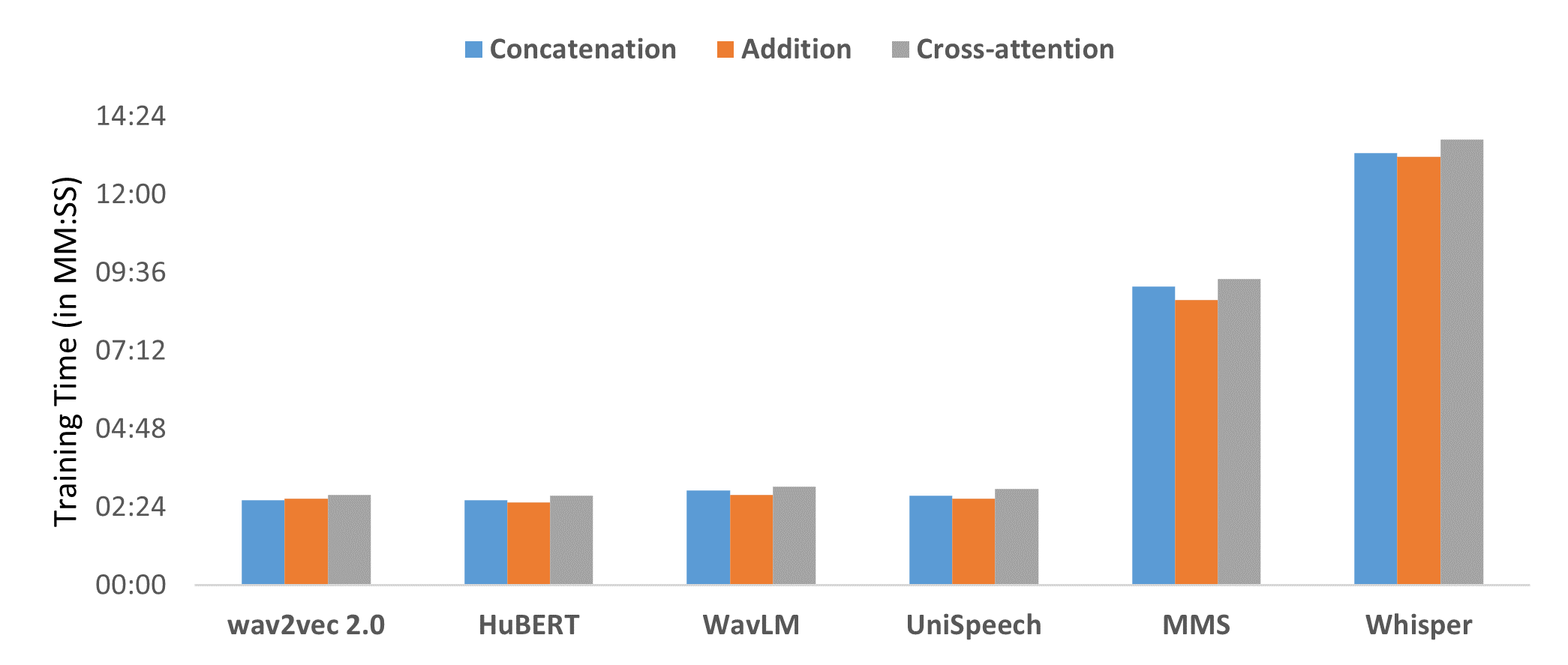}
    \vspace{-2mm}
    \caption{Analysis of training time for all the FusionVAD models using different fusion techniques (M: minutes and S: seconds).}
    \label{fig:time}
    \vspace{-4mm}
\end{figure}

\section{Results and Analysis}
\subsection{MFCC vs PTM Features}
First 3 columns in Table \ref{tab:der} shows the performance of VAD with individual features in terms of DER, FAR and MR. Whisper shows better performance than all other models. It is observed that MFCCs show high FAR than all other PTM features and lower MR than all PTM features except whisper. Also, it can be seen that all PTM features suffer from high MR than MFCC except Whisper. This pattern reveals that MFCC has information which helps in reducing MR, whereas PTM features has information which can reduce FAR. Through this pattern it can be hypothesized that MFCC might be detecting all high energy regions as speech including noisy, which results in high FAR. Whereas PTM features are correctly eliminating noise but also missing out many speech segments at the same time resulting in high MR. This shows that both features have complementary information, which can help to improve their worse counter parts if fused together.

\subsection{Comparison of Feature Fusion Techniques}

Table \ref{tab:der} presents the VAD performance using different feature fusion techniques applied to various PTM features. The results clearly indicate that all fusion-based models outperform their respective base models, demonstrating the effectiveness of feature fusion. The improvement in DER is mainly from a reduction in MR, which can be hypothesized because of MFCCs. This highlights the advantage of combining features that provide complementary information.


Additionally, cross-attention (CA) models consistently perform worse than concatenation and addition across all PTM features. This suggests that CA may not be optimal for this task. Among simpler fusion methods, addition outperforms concatenation in four out of six cases.
Table \ref{tab:pyannet-whisper} presents a dataset-wise comparison between the best fusion model, that is, fusion of MFCC and Whisper with addition and the SOTA Pyannote VAD. The fusion model consistently outperforms Pyannote across all three datasets, achieving an absolute average DER improvement of 2.04\%. To further validate our approach, we experimented with the rVAD method \cite{rvad} and used multi-resolution cochleagram (MRCG) features for comparison \cite{mrcg}. Our model outperformed rVAD by approximately 12\%. However, incorporating MRCG features led to a performance drop of around 2\% compared to using MFCC features alone in our best-performing model Whisper-MFCC-Addition.

Figure \ref{fig:fusion-comparison} presents the predictions obtained using different feature fusion techniques across various PTMs for a selected segment from the AMI corpus. The results indicate that concatenation and addition models produce boundaries that closely align with the ground truth, whereas CA fails to maintain this consistency in all cases. This trend aligns with the DER pattern observed in Table \ref{tab:der}, reinforcing that simpler fusion techniques outperform the more complex CA method for VAD. We hypothesize that VAD is inherently a simpler task that does not require extensive contextual information, unlike more complex tasks such as ASR. Additionally, the cross-attention method introduces a higher number of trainable parameters, as evident from Figure \ref{fig:parameters}. Due to this increased complexity, training time is also slightly longer for cross-attention compared to concatenation and addition, as shown in Figure \ref{fig:time}.

Among the three fusion techniques, CA consistently demands the most computational resources, both in terms of trainable parameters and training time. Concatenation, while slightly heavier than addition, remains significantly more efficient than CA. Cross-attention requires up to ~10\% more training time than addition, making it the most computationally expensive fusion approach, while addition remains the most parameter-efficient option.

\section{Conclusion}
This study investigates the impact of different feature fusion techniques for Voice Activity Detection (VAD) by combining hand-crafted MFCC features with pre-trained model (PTM) features. Our experiments show that simple fusion methods like addition and concatenation consistently outperform the more complex cross-attention mechanism. The results indicate that VAD, being a relatively simple task, does not benefit from attention-based feature fusion, which adds unnecessary computational overhead. Addition emerges as the most effective fusion strategy in four out of six models, while concatenation also performs well. Furthermore, our best fusion model surpasses the state-of-the-art VAD model (Pyannote) with an absolute improvement of DER of 2. 04 \% across datasets. These findings highlight that incorporating complementary features using lightweight fusion techniques enhances VAD performance while maintaining efficiency. Future work can explore extending these insights to other speech processing tasks, where the balance between complexity and effectiveness remains crucial.


\bibliographystyle{IEEEtran}
\bibliography{mybib}

\end{document}